\documentclass{elsart4}

\newcommand{\figref}[1]{Fig.~\protect\ref{#1}}
\usepackage[latin1]{inputenc}

 \usepackage{graphicx}

\usepackage{amssymb}

\begin{document}

\begin{frontmatter}

\title{Frustration effects in magnetic molecules}

\author[aff1]{R.~Schmidt}
\ead{reimar.schmidt@phyisk.uni.magdeburg.de}
\author[aff1]{, J.~Richter}
\author[aff2]{ and J. Schnack}
\address[aff1]{Institut f\"ur Theoretische Physik, Universit\"at Magdeburg,
  P.O. Box 4120, D-39016 Magdeburg, Germany}
\address[aff2]{Universit\"at Osnabr\"uck, Fachbereich Physik, Barbarastr. 7,
  D-49069 Osnabr\"uck, Germany}




\begin{abstract}
 By means of
 exact diagonalization we study the ground-state and the low-temperature
 physics of the Heisenberg antiferromagnet on the cuboctahedron
 and the icosidodecahedron. Both are  frustrated magnetic polytopes  
and correspond to the arrangement of magnetic atoms in 
the magnetic molecules Cu$_{12}$La$_8$ and Mo$_{72}$Fe$_{30}$. 
The interplay of strong quantum fluctuations and frustration influences the
ground state spin correlations drastically and  leads to an interesting
magnetization process at low temperatures. Furthermore the frustration yields
low-lying non-magnetic excitations resulting in  an extra 
low-temperature peak in the specific heat.
\end{abstract}


\end{frontmatter}
\begin{figure}[ht!]
  \begin{center}
    \scalebox{0.36}{
      \includegraphics[clip]{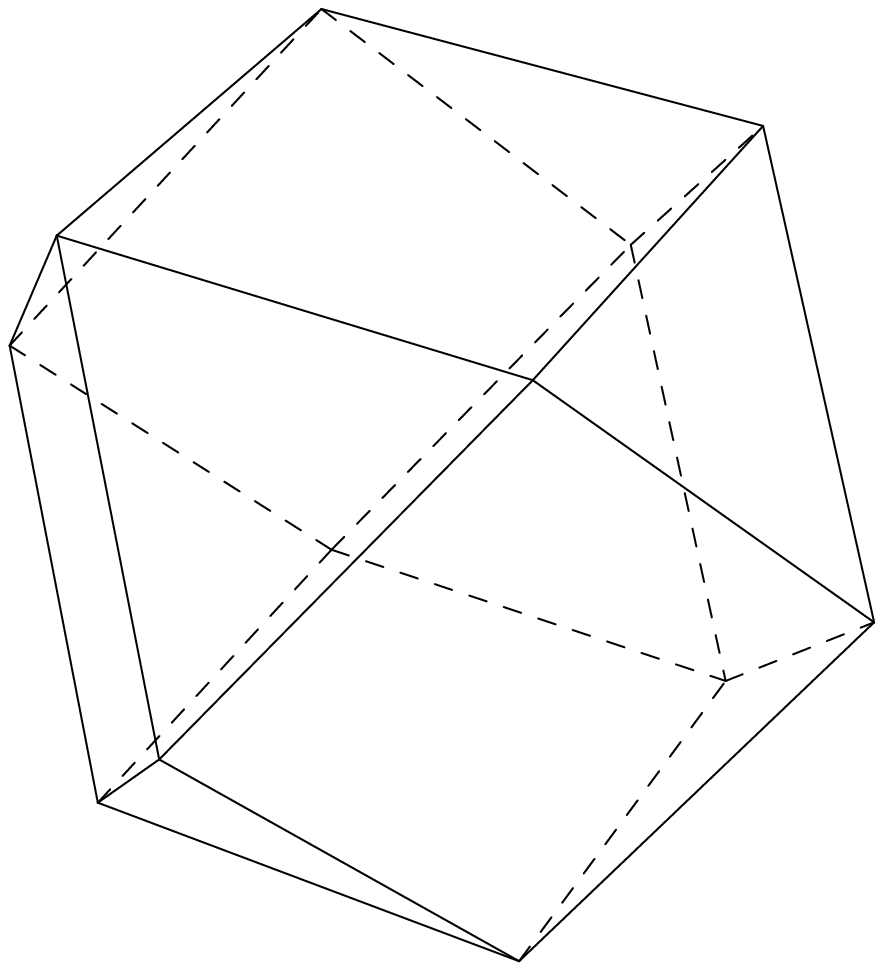}
      }
    \qquad
    \scalebox{0.28}{
      \includegraphics[clip]{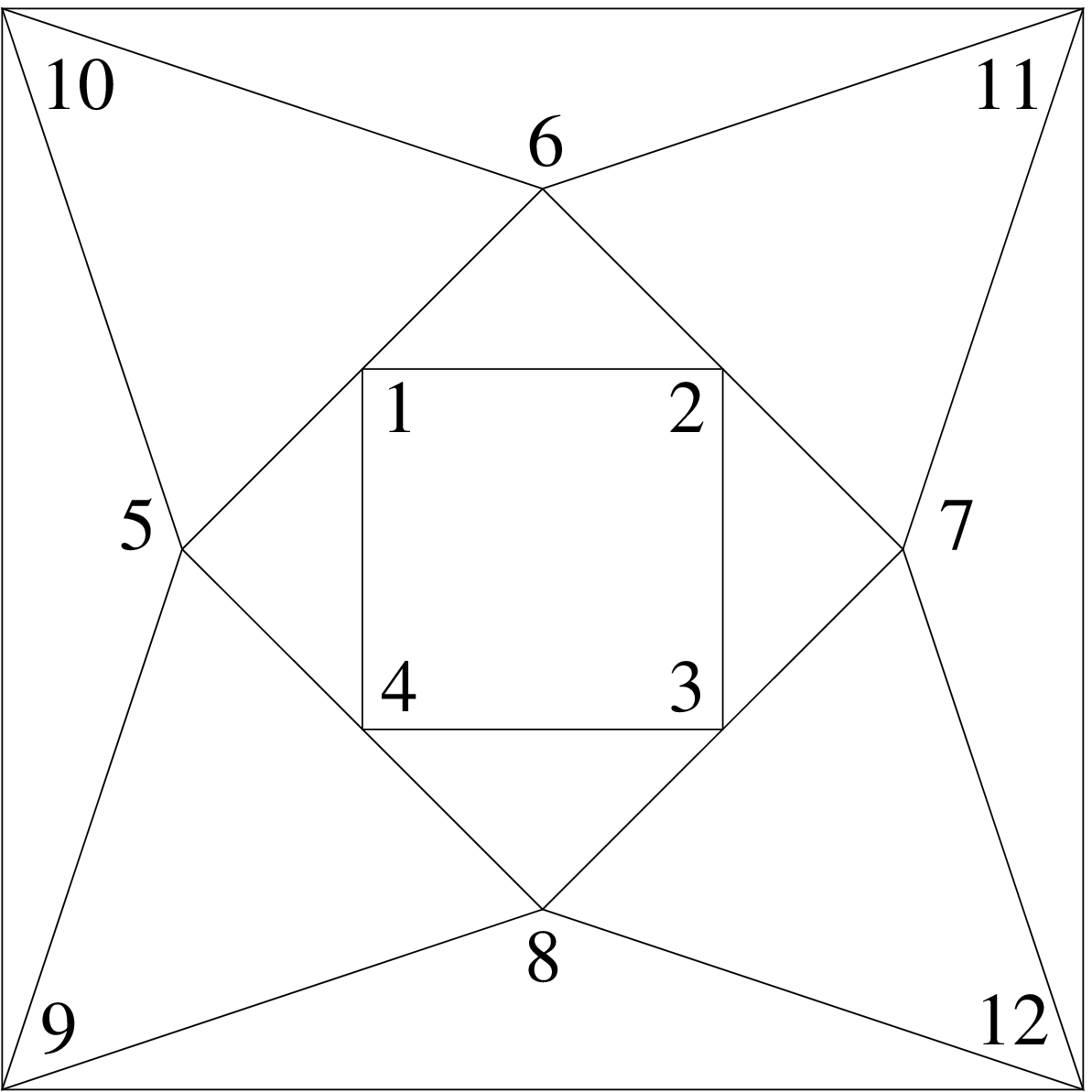}
      }
    \caption{\label{fcubo}The cuboctahedron  and its planar projection. 
      The spins sitting on the vertices are labeled by numbers. The 
      lines denote exchange couplings $J$.} 
  \end{center}
\end{figure}
\begin{figure}[ht!]
  \begin{center}
    \scalebox{0.32}{
      \includegraphics[clip]{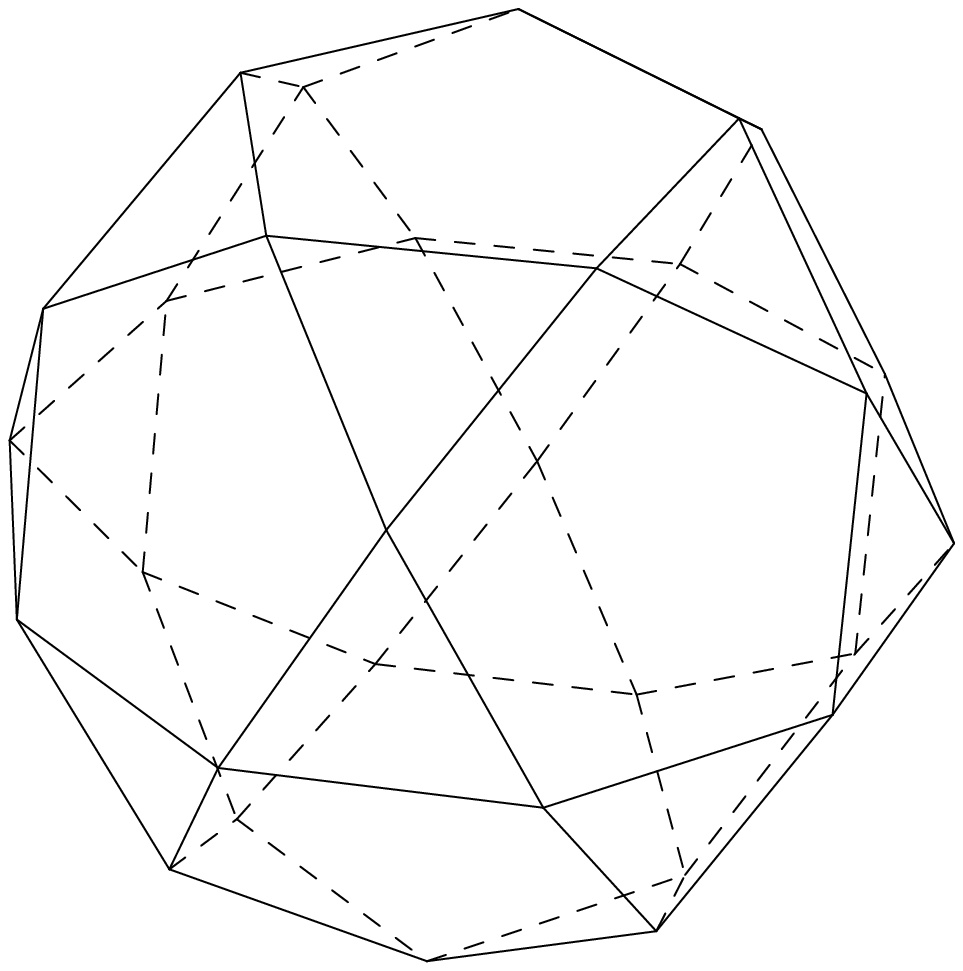}
     }
   \qquad
   \scalebox{0.15}{
      \includegraphics[clip]{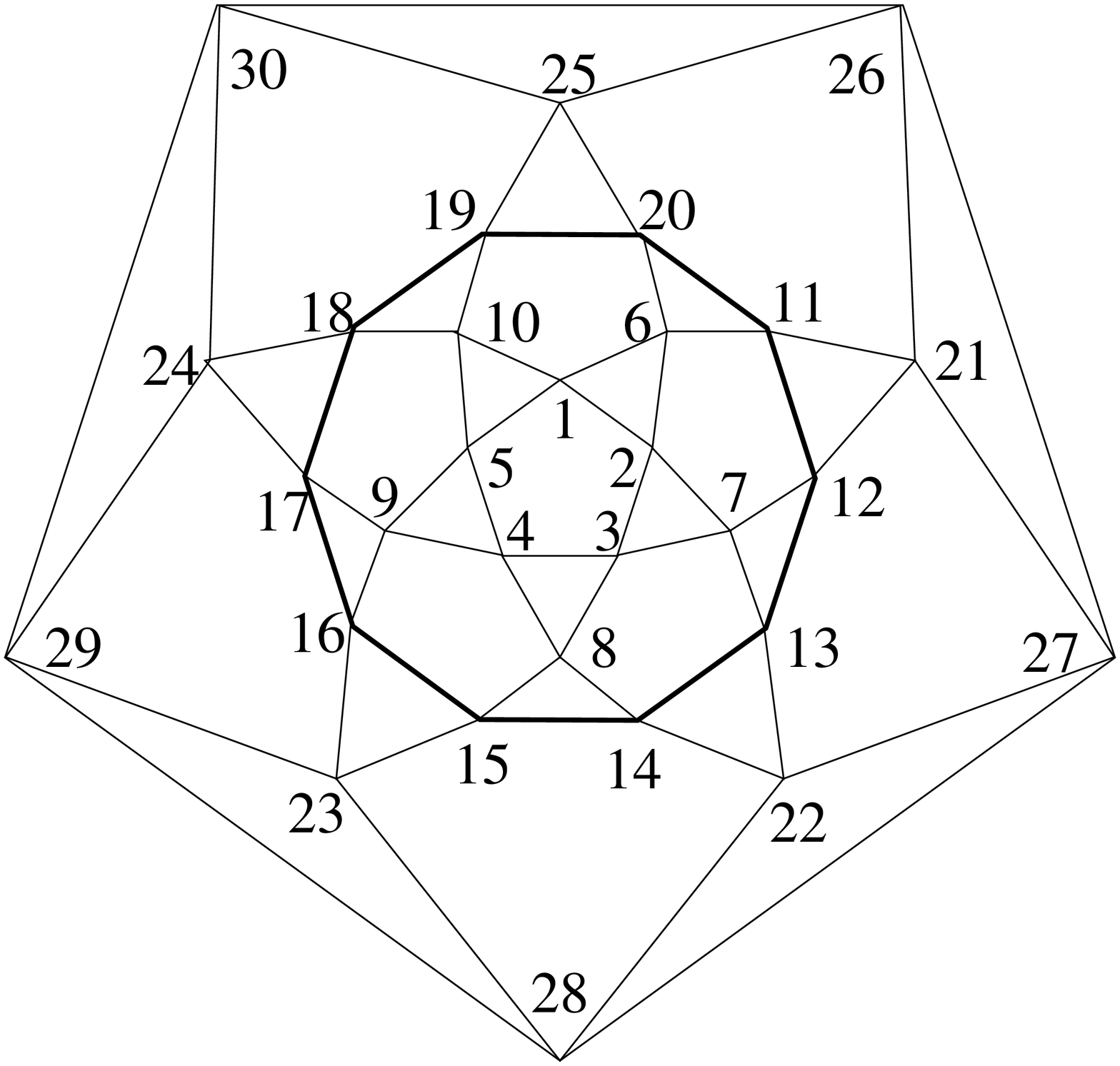}
      }
    \caption{\label{fico}
    The icosidodecahedron  and its planar projection. 
      The spins sitting on the vertices are labeled by numbers. The 
      lines denote exchange couplings $J$.}
  \end{center}
\end{figure}

{\it Introduction: $\;$} The interest in the properties of nanometer-sized
 magnetic molecules has 
 greatly advanced in recent years \cite{schnack04}. 
Besides spin rings of various lengths a new route to molecular magnetism is
based on highly frustrated magnetic polytopes like the cuboctahedron, 
Fig.~\ref{fcubo}, and  the icosidodecahedron, Fig.~\ref{fico}. Both are
 so-called 
Archimedean solids and magnetic frustration appears due to 
a triangular arrangement
of antiferromagnetic exchange bonds.
Recently magnetic molecules have been synthesized 
having spins located on the vertices of the cuboctahedron 
(Cu$_{12}$La$_8$ \cite{cubo}) and of
the icosidodecahedron (Mo$_{72}$Fe$_{30}$ \cite{acie38_3238,cpc2_517}).
In these frustrated magnetic systems the interplay of topology, interactions 
and  quantum fluctuations may lead to interesting physics at 
low temperatures $T$ as it is known from quantum spin lattices \cite{wir04}.

{\it The model: $\;$}
We consider a Heisenberg antiferromagnet
\begin{eqnarray}
\label{ham}
H
&=&
J\,
\sum_{(i<j)}\;
{\bf{s}}_i \cdot {\bf{s}}_j
+g \mu_B B_z S_z\ , \quad J > 0
\end{eqnarray}
to describe the magnetic properties. We set the energy scale by fixing the
exchange constant $J=1$.
The  ${\bf{s}}_i$ are the spin
operators of the individual magnetic
ions, 
${S}_z =\sum_i {s}^z_i$ is the $z$
component of the total spin ${\bf S}$ 
and $B_z$ is the magnetic field. We define 
$h=-g \mu_B B_z$ and in what follows we use $h$ as 
parameter of the field.
The strength of quantum fluctuations present in the model depends on 
the spin length ${\bf{s}}_i^2=s(s+1)$. While $s=1/2$ represents 
the extreme quantum case the system becomes classical in the limit $s \to
\infty$.  

\begin{figure}[ht!]
  \begin{center}
    \scalebox{0.62}{
      \includegraphics[clip]{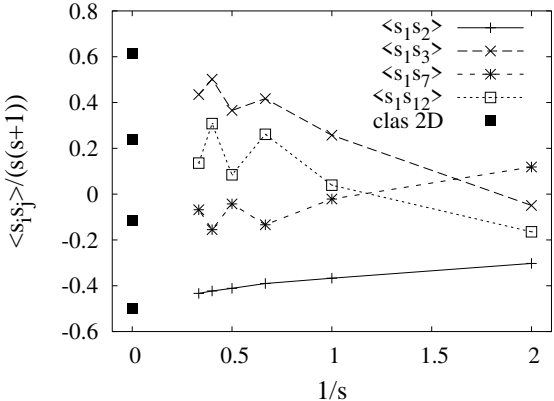}
      }
    \caption{\label{gssisj}The four different GS 
     correlators $\langle {\bf{s}}_i \cdot {\bf{s}}_j\rangle /s(s+1)$ of the 
     cuboctahedron 
     for $s=1/2,1,3/2,2,5/2,3$ versus  $1/s$.  
     For comparison we present the classical (i.e. $1/s=0$) 
     values obtained by averaging
       over the degenerate planar (clas) GS's.}
  \end{center}
\end{figure}

{\it The Cuboctahedron: $\;$}
The  
cuboctahedron is shown in Fig. \ref{fcubo}, left panel. To see the
arrangement of exchange bonds more clearly  we also show a planar projection
in  the right panel.
For this 12-spin system we are able to find the quantum ground state (GS) 
for
spin quantum numbers $s=1/2, 1, 3/2, 2, 5/2, 3$ and the full thermodynamics 
for
$s=1/2$. In Fig.\ref{gssisj} we show the different GS correlators in
dependence on the strength of quantum fluctuations measured by $1/s$.
Interestingly, the extreme quantum case $s=1/2$ differs from all other $s$
in the sign of the correlators except for the strong nearest-neighbor
correlator which is negative for all $s$. Further we see a qualitative
difference between half-integer and integer spin $s$. 
This comes from the fact that
the three spins on a triangle can be composed to a zero total spin
for  integer $s$, but cannot 
for half-integer $s$. 
The $1/s$ extrapolation of 
$\langle {\bf{s}}_i \cdot {\bf{s}}_j\rangle /s(s+1)$ approaches the
classical value obtained  taking into account only planar spin
configurations. 

\begin{figure}[ht!]
  \begin{center}
    \scalebox{0.62}{
      \includegraphics[clip]{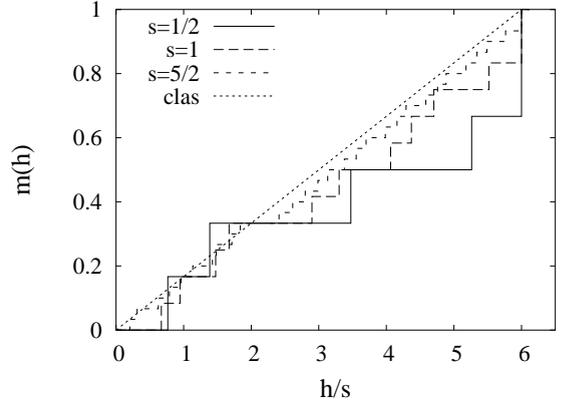}
      }
    \caption{\label{cubogsm}$T=0$ magnetization $m$ in dependence on 
$h/(Js)$ for the cuboctahedron with
      $s=1/2,1,5/2$ and in the classical case ($s\to \infty$).}  
  \end{center}
\end{figure}

Next we consider the magnetization $m= {S}_z /(Ns)$ 
versus field $h$ curve  
at zero temperature 
shown in Fig.\ref{cubogsm}. 
In the classical limit it is a straight line up
to saturation field $h_s/s=6$. The $m(h)$ curve for quantum model with $s=1/2$ 
differs drastically from the classical straight line. Though
for any finite spin system one has step-like $m(h)$ 
curves there are two special
features for the cuboctahedron to be mentioned.
The first one is the height of the last step to saturation $m=1$ which is
twice as large as the other steps. Again this is a frustration effect and is
related to so-called localized magnon states trapped on squares
\cite{schnack04,wir04,prl88_167207}.  The second interesting feature is 
the large width of the plateaus for $m=1/3$ and $m=1/2$. The existence of
plateaus  caused
by the interplay of frustration and quantum fluctuations  has been
recently widely discussed for spin lattices \cite{hon04,wir04}.  
Both features are less pronounced also
present for $s=1$ but vanish for larger $s$.  

\begin{figure}[ht!]
  \begin{center}
    \scalebox{0.62}{
      \includegraphics[clip]{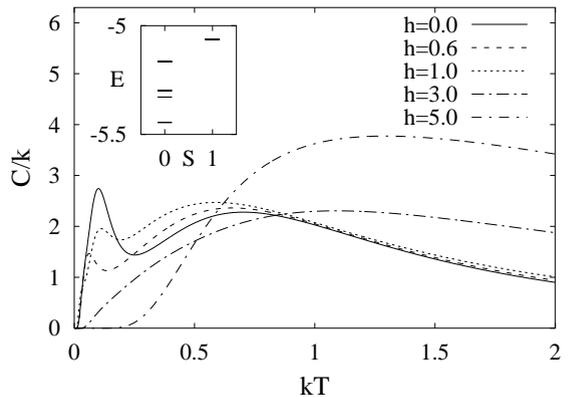}
      }
    \caption{\label{cuboc}Specific heat of the cuboctahedron with
      $s=1/2$ versus temperature $T$ for different magnetic fields $h$. The
      inset shows the lowest energies for $S=0,1$. Note that the third singlet
      excitation is twofold and the fourth singlet 
      excitation is threefold degenerated.}  
  \end{center}
\end{figure}

Let us now briefly discuss the specific heat $C$ 
for the cuboctahedron with
$s=1/2$ for various strengths of magnetic field $h$ in \figref{cuboc}.
The most striking feature is the low-$T$ peak in $C$ in zero and low
fields. The peak disappears for larger fields. We mention, that spin rings do
not show such an additional low-$T$ peak
in the zero-field specific heat.
The existence of such a peak is related to the low-lying spectrum of the
system. Contrary to unfrustrated spin rings we find for the 
cuboctahedron with $s=1/2$ at $h=0$ 
seven low-lying singlet excitations below the first triplet excitation which
might be important for the extra peak (see the inset in \figref{cuboc}).

{\it The icosidodecahedron: $\;$}
Now we consider the icosidodecahedron, Fig. \ref{fico}, left panel. 
To see the
arrangement of exchange bonds more clearly  we also show a 
planar projection
in the right panel.
For this 30-spin system we are able to find the quantum GS 
at arbitrary magnetic field $h$ for
spin quantum numbers $s=1/2$ only. 
For larger spin $s$ we are able to calculate the GS for 
large fields only.
The GS correlators for $s=1/2$ at zero field are 
 $\langle {\bf s}_1 {\bf s}_2 \rangle = -0.22057$,
 $\langle {\bf s}_1 {\bf s}_3 \rangle =-0.00026$,
 $\langle {\bf s}_1 {\bf s}_7 \rangle = 0.05915$,
 $\langle {\bf s}_1 {\bf s}_8 \rangle = 0.00011$,
 $\langle {\bf s}_1 {\bf s}_{13} \rangle = -0.03609$, 
 $\langle {\bf s}_1 {\bf s}_{14} \rangle = -0.00184$,
 $\langle {\bf s}_1 {\bf s}_{22} \rangle = 0.02097$,
 $\langle {\bf s}_1 {\bf s}_{28} \rangle = -0.03592$.
Obviously, all correlations except the nearest-neighbor one are weak as a
result of frustration and quantum fluctuations. 

\begin{figure}[ht!]
  \begin{center}
    \scalebox{0.62}{
      \includegraphics[clip]{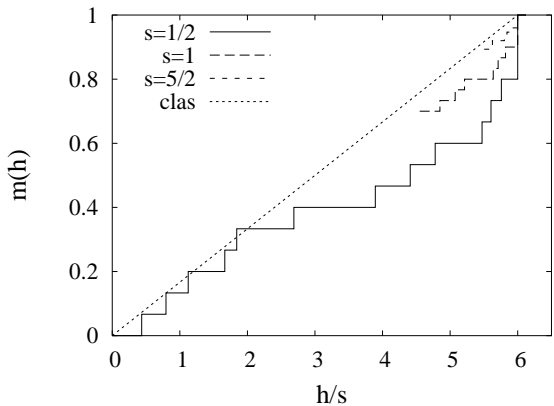}
      }
    \caption{\label{icogsm}$T=0$ magnetization $m(h)$ in dependence on 
$h/(Js)$
for the icosidodecahedron with
      $s=1/2,1,5/2$ and in
      the classical case ($s\to \infty$).}   
  \end{center}
\end{figure}

In Fig.\ref{icogsm}
we show the magnetization $m= {S}_z /(Ns)$ 
versus field $h$ curve  
at $T=0$. 
Again for $s\to \infty$ it is a straight line up
to saturation field, but for $s=1/2$ it is
completely different. Like for the cuboctahedron the $m(h)$ curve
for $s=1/2$ 
shows an extra high jump to saturation due to localized magnon states
and plateaus of large width at $m=3/5, 2/5, 1/3$. 
For $s=1$ and $s=5/2$ we show the high-field part of $m(h)$. The jump and
the plateaus are also seen for $s=1$, but less pronounced. For $s=5/2$ the
$m(h)$ curve approaches the classical line. 

\begin{figure}[ht!]
  \begin{center}
    \scalebox{0.62}{
      \includegraphics[clip]{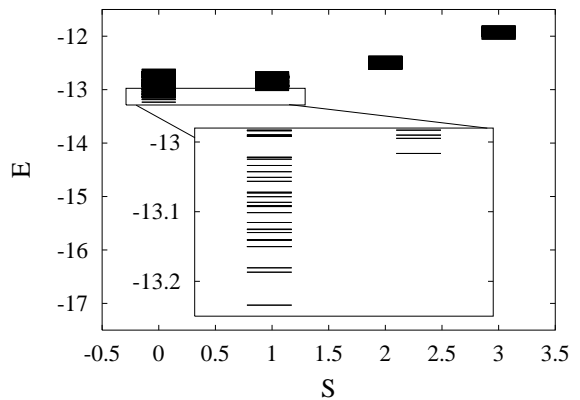}
      }
    \caption{\label{icoll}Low-lying levels versus quantum number of total
     spin $S$ for the icosidodecahedron with $s=1/2$. 
      The inset shows the singlets ($S=0$) below the
      first triplet ($S=1$) on an enlarged scale.}  
  \end{center}
\end{figure}

Though we
cannot calculate thermodynamic quantities we are able to present
the low-energy spectrum, relevant for the low-$T$ physics,
 for the icosidodecahron with $s=1/2$ (\figref{icoll}). 
The lowest states in each sector of the total spin $S$
are well described by rotational bands, i.e.  
we have approximately $E_{min}(s) \propto S(S+1)$ \cite {schnack04}. 
As shown in  the inset there are more than 20 singlets ($S=0$) below the
first triplet ($S=1$) excitation. This large number of  low-lying 
singlets
is due to frustration and was also found for the cuboctahedron with
$s=1/2$. Note that unfrustrated rings do not have singlet excitations below the
first triplet. Hence we may expect that for the icosidodecahedron also an
extra low-$T$ peak in $C$ appears.
We mention, however, that for the low-$T$ behavior of the susceptibility
the singlets are irrelevant.

{\it Acknowldegement:} We are indebted  to J. Schulenburg for 
assistance in numerical calculations.


\end{document}